\documentclass[twocolumn,aps,prl,preprintnumbers,amsmath,amssymb,showpacs]{revtex4-1}

\usepackage{amsmath}                                
\usepackage{amssymb}
\usepackage{amsxtra}
\usepackage{amsfonts}
\usepackage{color}
\usepackage[latin1]{inputenc}                        
\usepackage[dvips]{graphicx}                         

\begin{document}

\title{Spin-orbit coupling effect in (Ga,Mn)As films: anisotropic exchange
  interactions and magnetocrystalline anisotropy}

\author{S.~Mankovsky$^1$}
\author{S.~Polesya$^1$}
\author{S.~Bornemann$^1$}
\author{J.~Min\'ar$^1$}
\author{F.~Hoffmann$^2$}
\author{C.~H.~Back$^2$}
\author{H.~Ebert$^1$}
\affiliation{%
$^1$Department of Chemistry/Phys. Chemistry, LMU Munich,
Butenandtstrasse 11, D-81377 Munich, Germany
}  %
\affiliation{$^2$ Department of Physics, Universit{\"a}t Regensburg, 93040 Regensburg, Germany}


\begin{abstract}

The magneto-crystalline anisotropy (MCA) of (Ga,Mn)As films has been studied on the basis of
ab-initio electronic structure theory by performing magnetic torque calculations. An
appreciable contribution to the in-plane uniaxial anisotropy can be attributed
to an extended region adjacent to the surface. 
Calculations of the exchange tensor allow to ascribe a significant
part to the MCA to the exchange anisotropy, caused
either by a tetragonal distortion of the lattice  or by the presence of the
surface or interface.

\end{abstract}

\maketitle

Diluted  magnetic semiconductors (DMS) are a class of materials having
attractive properties for spintronic applications (e.g. see
review~\cite{JSM+06}).  Many investigations in this field are focussed on the
(Ga,Mn)As DMS system with 1 to 10\% of Mn atoms which have promising features
from a physical as well as technological point of view.  The crucial role of
valence states with respect to various magnetic properties of (Ga,Mn)As was
discussed in the literature by many authors~\cite{AJBM01,JSM+06}.  First of
all, the valence band  holes are responsible for ferromagnetic (FM) order in
the system mediating the exchange interaction between well localized Mn
magnetic moments.   Spin-orbit coupling of the states at the top of valence
band, being close to the Fermi level, leads to a rather strong cubic
magnetocrystalline anisotropy (MCA) in bulk (Ga,Mn)As and to an
in-plane biaxial MCA in the (Ga,Mn)As film on top of a GaAs substrate ~\cite{DOM01}. In the latter case the 
spin-orbit coupling (SOC) makes the valence states close to $E_F$ sensitive to
lattice distortions and in that way responsible for the in-plane MCA due to
compressive strains originating from the lattice mismatch between the (Ga,Mn)As
film and GaAs
substrate~\cite{Sav04,Die01,Sav03,LSF03,HWT+06,SCL+07,WVL+03,TKAR03,WSE+05,SKT+08,HVF+02}.
As soon as the spin polarization of the valence bands is rather small,
the MCA
in (Ga,Mn)As is discussed in terms of anisotropic exchange interactions of the
Mn atoms~\cite{DOM01,Sav04,AJBM01}.  The strength of the MCA depends on the
hole concentration introduced by the Mn impurity atoms~\cite{DOM01,HWT+06,SWE+05,GDD+09} as well as
on the variation of equilibrium lattice parameter of (Ga,Mn)As increasing with
the increase of Mn content and resulting in a larger lattice mismatch with the
GaAs substrate.

Numerous experimental results evidenced a transition from the bi-axial to the
uni-axial in-plane
anisotropy~\cite{HWT+06,SCL+07,WVL+03,TKAR03,WSE+05,SKT+08,HVF+02,SWE+05}.  So far,
however, there is no consensus in the literature concerning the origin of the
in-plane uniaxial anisotropy.  Although in some recent theoretical
works the origin of the uniaxial in-plane anisotropy is attributed to a trigonal
distortion caused by a uniaxial or shear strain  within the film
plane~\cite{SWE+05,WVL+03,ZKOJ09}, this type of distortion was not observed
experimentally.  
Stacking fault defects in the $(111)$ and $(11\overline{1})$ planes
have been found recently in experiment  
\cite{KKM+11} which could be responsible for breaking the
equivalence of the $[110]$ and $[1\overline{1}0]$ directions in the (Ga,Mn)As films. 
However, there is so far no experimental evidence nor theoretical
description showing that these stacking faults are responsible for the in-plane
uniaxial anisotropy.


In order to obtain a more detailed understanding of the subtle electronic
effects which determine the MCA properties of (Ga,Mn)As films, we performed
ab-initio electronic structure calculations for tetragonally distorted
(Ga,Mn)As bulk as well as (Ga,Mn)As films deposited on a GaAs substrate.
The ab-initio calculations have been performed within the framework of the
local spin density approximation (LSDA) of density functional theory (DFT)
using the fully relativistic Korringa-Kohn-Rostoker (KKR) multiple scattering
band structure method~\cite{Ebe00,SPR-KKR5.4}.  For the treatment of the
chemical disorder in (Ga,Mn)As alloys we applied the coherent potential
approximation (CPA).  Moreover, for the bulk and surface calculations we used a
regular $\vec{k}$-mesh of $63 \times 63 \times 63$ points in the full 3D
Brillouin Zone (BZ) and $63 \times 63$ points in the full 2D BZ, respectively.
For the angular momentum
expansion of the Green's function a cutoff of $\ell_{\rm max} = 3$ was applied.

\begin{figure}
  \begin{center} 
  \includegraphics[scale=0.2]{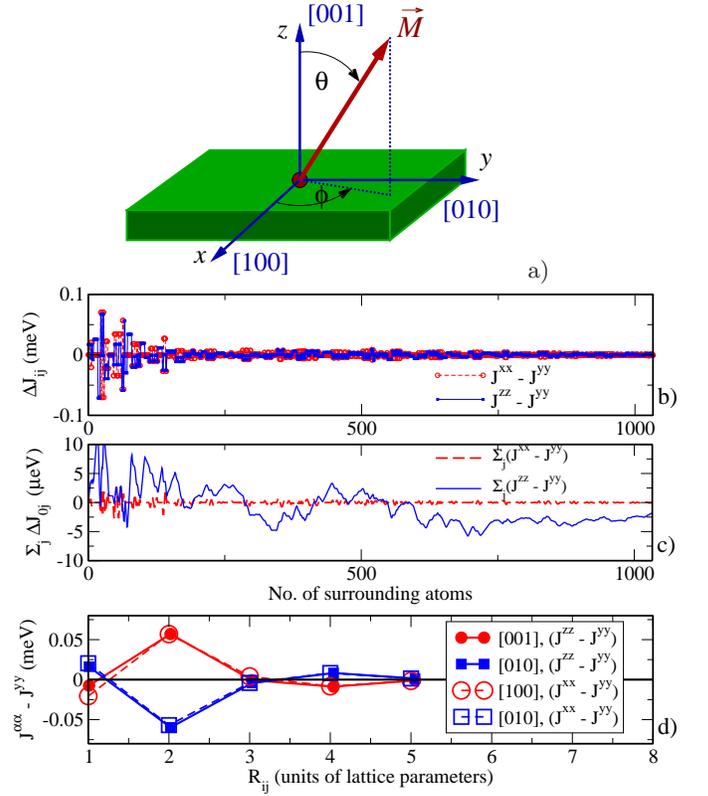} \;\;a)
  \includegraphics[scale=0.35]{J_zz_minus_yy_integral_2.eps} \;\; 
  \end{center}
  \vskip-3mm
  \caption{ \label{J_bulk}   a) Geometry for the torque calculations;
b) $J^{zz} - J^{yy}$ and  $J^{xx} - J^{yy}$  for bulk (Ga,Mn)As with 5\% Mn, with tetragonal distortion c/a = 1.01; c)  $\sum_j (J^{zz}_{ij} - J^{yy}_{ij})$ and  $\sum_j (J^{xx}_{ij} - J^{yy}_{ij})$ over all lattice sites up to $R_{ij} \leq 5a$; d) $J^{xx} - J^{yy}$  along [100] and  [010] directions. }
  \vskip-3mm 
\end{figure}

The study of the magneto-crystalline anisotropy (MCA) was performed by calculating the magnetic
torque  $\vec{T}_i^{(\hat{e}_i)}=-\partial
E(\{\hat{e}_k\})/\partial \hat{e}_i \times \hat{e}_i$ acting on the magnetic
moment $\vec{m}_i$ of the atomic site $i$, with a unit vector $\hat{e}_i =
\vec{m}_i/|\vec{m}_i|$ pointing along the direction of the magnetization
$\vec{M}$. The 
component of the magnetic torque with respect to the axis $\hat{u}$
\begin{eqnarray}
T_{\hat{u}} (\theta, \phi) &=& -\partial E(\vec{M}(\theta,
\phi))/\partial  \theta
\end{eqnarray}
was calculated from first-principles as described in \cite{SSB+06}. 
Here, the  $\hat{u}$ vector specified by the
angles $\theta$ and $\phi$ (see Fig. \ref{J_bulk}a) lies within the surface plane and is perpendicular
to the direction of the magnetic moment $\hat{e}_M$. For an uniaxial anisotropy
a special geometry can be used which gives a simple relationship between the
magnetic torque and the energy difference between the in-plane and out-of-plane
magnetization directions. Setting $\theta = \pi/4$, the torque component
$T_{\hat{u}}$ gives the $\phi$ dependent energy difference   $T_{\hat{u}}
(\theta = \pi/4, \phi) = E_{||} (\phi) - E_{\perp}$. In the case of an in-plane
anisotropy these values can also be used to evaluate the anisotropy energy
within the plane, comparing in particular
the directions $[110]$ and $[1\overline{1}0]$.

The exchange coupling tensor $\underline{\underline{J}}_{ij}$
used below for the discussions of the magnetic anisotropy 
in terms of the anisotropic Mn-Mn exchange interactions~\cite{AJBM01,DOM01,Sav04}
was calculated as described in Ref.~\cite{EM09a}. 
Here, the effective coefficients of the uniaxial MCA are represented  
by the following expression~\cite{USPW03,MBM+09}:
\begin{eqnarray} \label{K_effective}
\tilde{K}_i &=& -\sum_j (J^{zz}_{ij} - J^{xx}_{ij}) + 2{K}_i\;, 
\end{eqnarray}
with ${K}_i$ being the on-site MCA coefficients~\cite{USPW03}.

In order to study the strain-induced effect in the MCA of deposited (Ga,Mn)As
films, we consider at first a bulk system with tetragonal distortion
(avoiding surface and interface contributions) which is then
characterized by the $c/a$ ratio.  Magnetic torque calculations simulating the
strain-induced effects in the alloy with 5\% Mn yield a linear variation of the
magnetic anisotropy energy, $E_{[100]} - E_{[001]}$,  from $+3.38$ to
$-3.37\;\mu$eV per unit cell for $c/a$ ratio varying from 0.99 to 1.01, i.e.
the magnetic easy axis changes from an out-of-plane to an in-plane orientation,
which is in line with corresponding experimental data \cite{DSG+08}. 
As the $[100]$ and $[010]$ directions are equivalent, this leads to the
bi-axial in-plane MCA with $[100]$ and $[010]$ being easy
magnetization directions and an in-plane anisotropy energy 
$E_{[100]}- E_{[110]} \approx -0.1$ \;$\mu$eV per unit cell.

For a more detailed analysis of the relationship between the MCA and anisotropy
of Mn-Mn exchange interactions, calculations of the exchange coupling tensor elements $J^{\alpha\beta}_{ij}$
have been performed for (Ga,Mn)As 
with 5\% Mn both without any distortion as well as with a tetragonal distortion of $c/a = 1.01$.
For an undistorted (Ga,Mn)As system we find that the sum
$\sum_j (J^{\alpha\alpha}_{ij} - J^{\beta\beta}_{ij})$ ($\alpha,\beta = x,y,z$) over
 all lattice sites in the expression given in Eq.~(\ref{K_effective}) vanishes
 (see Fig. \ref{J_bulk}b).
This is a consequence of the system's  symmetry,
in spite of the fact that the individual terms $(J^{\alpha\alpha}_{ij} -
J^{\beta\beta}_{ij})$ with $\alpha \neq \beta$  are non-zero.
In the presence of a tetragonal distortion along the $z$-axis, the symmetry
properties within the $xy$ plane (i.e. (001) plane) do not change. Therefore,
summation over all lattice sites up to $R_{ij} = 5a$ (with lattice parameter $a$)
shown in Fig.~\ref{J_bulk}c gives $\sum_j (J^{xx}_{ij} - J^{yy}_{ij}) = 0$.
For more details,  Fig. \ref{J_bulk}d shows the differences  $J^{xx}_{ij} -
J^{yy}_{ij}$ for $\vec{R}_{ij}$ taken along [100] and  [010] directions
(dashed lines).  These values are finite and equal in magnitude,
but they have an opposite sign and therefore cancel each other upon summation
over all sites.  
However, due to the tetragonal distortion along $z$, $(J^{zz}_{ij} - J^{yy}_{ij})$
for $\vec{R}_{ij}$ taken along [001] and [010] directions are not equivalent
(see Fig. \ref{J_bulk}d, solid lines) and thus the sum  $\sum_j (J^{zz}_{ij} - J^{yy}_{ij})$
over all lattice sites does not vanish anymore.  The summation over all lattice
sites up to $R_{ij} \leq 5a$  is shown in Fig.~\ref{J_bulk}c which gives the
contribution to the uniaxial MCA that originates from the exchange anisotropy
being $\approx 2.5\;\mu$eV.  Because of the slow convergence of the sum with
increasing distance, this gives only an
approximation to the true contribution due to the exchange
anisotropy. Nevertheless, the value 
obtained in this way has the same order of magnitude as the MAE obtained from
our torque calculations leading to the conclusion that the exchange anisotropy
has indeed a significant impact on the total MAE.


Our present investigations of the in-plane uniaxial anisotropy have been
performed for a 8 monolayer (ML) thick (Ga,Mn)As film deposited on a 
semi-infinite (001)-oriented GaAs substrate.  In order to distinguish the anisotropy behaviour in
the vicinity to the interface with GaAs as well as in the area adjacent to the
surface we performed calculations for an uncovered (Ga,Mn)As film as well as one
with two additional capping layers of Au.  Due to small amount of free charge
carriers in (Ga,Mn)As the surface potential decays slowly into bulk leading to
a potential and a charge density gradient within an extended region adjacent to
the surface. The existence of such a potential gradient results in the breaking
of the 4-fold symmetry of the bulk (Ga,Mn)As system, making
the $[1\overline{1}0]$ and $[110]$ directions inequivalent (for the
geometry used here this corresponds to the $x$ and $y$ directions,
respectively) and leading effectively to a $C_{2v}$ symmetry not only within the few
surface/interface layers but also in a rather extended subsurface
regime. 

We discuss now the surface induced MCA in the film.
Here we focus mainly on the MAE properties of a (Ga,Mn)As film with a clean Ga
terminated surface deposited on GaAs(001).   The results for the energy
differences between different magnetization directions are $E_{[110]} -
E_{[001]} = -80.56\; \mu$eV and  $E_{[1\overline{1}0]} - E_{[001]} =
-32.96\;\mu$eV per film unit cell (8~ML).  This gives an uniaxial in-plane
anisotropy with the energy difference of $E_{[110]} - E_{[1\overline{1}0]} =
-47.6\; \mu$eV per film unit cell. 

\begin{figure}
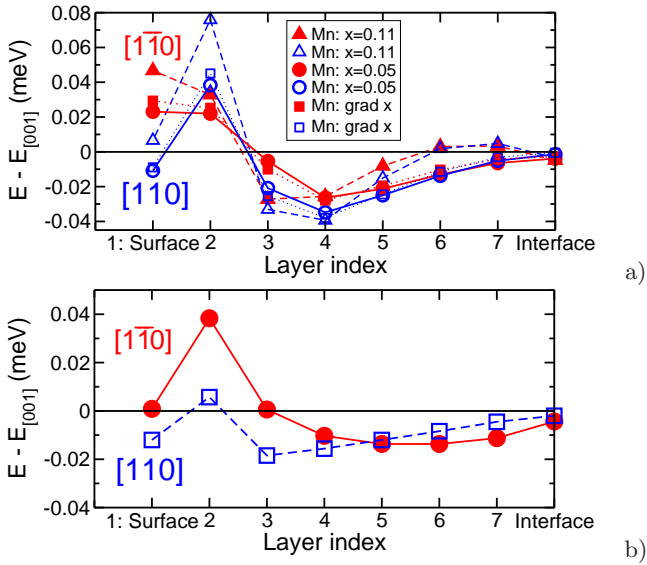

  \begin{center}
  \includegraphics[scale=0.3]{CMP_TORQUE_vs_layer_Mn0.05_vs_grad_8ML_T_7.1.eps} \;\; a)
  \includegraphics[scale=0.3]{CMP_TORQUE_vs_layer_Mn0.05_Al.eps} \;\; b)
  \end{center}
  \vskip-3mm
  \caption{a) Layer resolved
    contributions to the MCA 
    energy in the uncovered 8ML (Ga,Mn)As film with 5 at.\% Mn (circles) and  11(triangles) 
    at.\% Mn, as well as with Mn content varying from 5 at.\% at the
    interface to 6.6 at.\% in surface layer ($grad x$, squares), for two directions of magnetization:
    $\vec{M} || [110]$ and  $\vec{M} || [1\overline{1}0]$; b) layer resolved
    contributions to the MCA 
    energy in the 8ML (Ga,Mn)As film with 5 at.\% Mn, with 3 capping layers of Au.
   }
  \vskip-3mm
  \label{torque}
\end{figure}

Fig. \ref{torque}a presents the layer resolved contributions to the $E_{[110]}
- E_{[001]}$ and $E_{[1\overline{1}0]} - E_{[001]}$ values, indicated by open
  and filled symbols, respectively. The difference between these values
characterizes the MCA within the plane. One should emphasize here that the
contribution to the MCA from the region close to the surface decays slowly into
the bulk. Therefore the surface-induced anisotropy effect in the uniaxial
in-plane MCA is determined by a rather extended region adjacent to the surface
and not just by two or three subsurface layers as it is often
observed in metallic systems (e.g. \cite{Kos07}). The corresponding contribution to the  energy of
the uniaxial in-plane anisotropy exceeds by far the energy of the bi-axial in-plane
anisotropy when normalized to the same volume ($E_{[100]}- E_{[110]} \approx -0.1$
\;$\mu$eV per unit cell of the bulk system).    Using these results the MCA of
experimental (Ga,Mn)As films consisting of $n+8$ monolayers can be modelled by
combining the contribution of $n$ bulk-like layers with the contribution of 8
layers of surface region.  This gives two competing contributions to the MCA: a
bi-axial in-plane anisotropy from bulk-like layers of (Ga,Mn)As with a
tetragonal distortion and a uniaxial in-plane anisotropy from the area adjacent
to the surface.  Applying our obtained MAE values to a unit volume, one can get
the MCA of the whole film including the surface region. Within our
consideration, the coefficient of the in-plane bi-axial anisotropy $K_4$ does
not depend on the film thickness $L$, while the the coefficient of the
in-plane uni-axial anisotropy  $K^{||}_2$ recalculated per unit volume should
decrease with film thickness as $1/L$.  Thus, according to our numerical
results, a rather strong uniaxial anisotropy should be observed in the case of
very thin films, while the increase of the film thickness should lead to a
competition of bi-axial and uni-axial anisotropies beginning with a certain
film thickness. 
The contribution from the 'surface' region to the out-of-plane uniaxial
anisotropy  $E_{[1\overline{1}0]} - E_{[001]}$ decreases as well with the
film thickness as  $1/L$. This results in a leading role
of the in-plane anisotropy contribution caused by 
the tetragonal lattice distortion discussed above.
Note that an increase of the Mn concentration results in an
increase of the charge carriers in the film which again results in better
screening of the surface potential. This can be seen in Fig. \ref{torque}a,
where the values $E_{[110]} - E_{[001]}$ and $E_{[1\overline{1}0]} - E_{[001]}$
are shown as a function of the distance from the surface for a (Ga,Mn)As film with
11\% Mn. This increase in Mn concentration results in an in-plane MAE
$E_{[110]} - E_{[1\overline{1}0]} = -20.8\; \mu$eV per film unit cell, i.e. one
obtains a smaller anisotropy energy when compared to the case of 5\%Mn.

Since the uniaxial MCA has its origin in an extended subsurface region one can
expect that it is an intrinsic property of the systems and should be observed
not only in the case of a clean surface but also in the presence of overlayers
on the top of the (Ga,Mn)As film.  Corresponding investigations have been
performed for a (Ga,Mn)As film with 3 capping layers of Au on top of the
(Ga,Mn)As film. The resulting layer resolved contribution to the MCA is shown in
Fig.~\ref{torque}b. In spite of the differences in the MAE between the Au capped
(Ga,Mn)As film and the case of uncovered film, the general trend in both cases is
the same, i.e. one can clearly see that the difference in layer contributions to
the MCE for different directions of magnetization, along $[110]$ and
$[1\overline{1}0]$, decays slowly with the distance from the surface or
Au/(Ga,Mn)As interface, respectively. 

To investigate also the effect caused by a concentration gradient
along the surface normal within an uncovered (Ga,Mn)As film we dealt with a
corresponding film where the Mn concentration varies from 5\% at the
(Ga,Mn)As/GaAs interface to 6.6\% in the surface layer. As can be seen in Fig.
\ref{torque}a such a gradient does not result in a noteworthy change in the
MCA. 

\begin{figure}
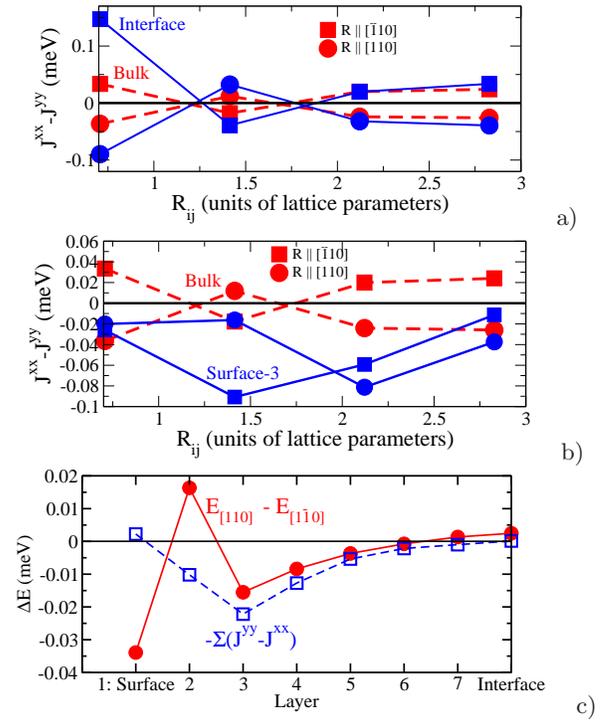

  \begin{center}
  \includegraphics[scale=0.33]{CMP_Jij_aniso_spin_interf.eps} \;\; a)\\
  \includegraphics[scale=0.33]{CMP_Jij_aniso_spin_surf_minus_3.eps} \;\; b)\\
  \includegraphics[scale=0.28]{CMP_Mn_0.05_Torque_vs_JxJy.2.eps} \;\; c)
  \end{center}
  \vskip-3mm
  \caption { Variation of $(J^{xx}_{ij} - J^{yy}_{ij})$
    with distance $R_{ij}$ of pairs $(i,j)$ of Mn atoms taken in
    $[1{1}0]$ and $[1\overline{1}0]$ 
    directions in the (Ga,Mn)As/GaAs film: a) bulk vs (Ga,Mn)As/GaAs
    interface and b) bulk vs (Surf.- 3)-layer ;
  c) Layer-resolved sum  $-\sum_{j} (J^{yy}_{ij} - J^{xx}_{ij})$ calculated within the
sphere of radius $2.9a$ in
comparison with the MCA energy $E^{[1{1}0]} -
E^{[1\overline{1}0]}$ evaluated by magnetic torque calculations for the
8ML (Ga,Mn)As film with 5 at.\%  Mn. 
}
  \vskip-3mm
  \label{DOS}
\end{figure}

To analyze in more detail the origin of the surface-induced in-plane uniaxial
anisotropy the contribution of the exchange interaction anisotropy in the
(Ga,Mn)As film  was determined.  Fig.~\ref{DOS} shows the difference
$J^{xx}_{ij} - J^{yy}_{ij}$ calculated along the $[1{1}0]$ and $[1\overline{1}0]$
directions within the film layers where the $x$ and $y$ axes are chosen along
$[1\overline{1}0]$ and $[110]$ directions, respectively.  As discussed above,
for bulk (Ga,Mn)As the variation of $J^{xx}_{ij} - J^{yy}_{ij}$ with
distance  $|\vec{R}_{ij}|$  is the same for $\vec{R}_{ij}$ along the $[1{1}0]$ and
$[1\overline{1}0]$ however with different sign. This behaviour is more
or less the same for Mn atoms next to the (Ga,Mn)As/GaAs interface (see
Fig.~\ref{DOS}a). For Mn in the fourth layer (chosing the surface layer
as the first layer), however, the situation is changed indicating a pronounced
modification of the anisotropic exchange coupling due to the broken symmetry.
As a result, the sum over lattice sites in Eq.~(\ref{K_effective}) does not
vanish which leads to a contribution to the in-plane uniaxial anisotropy.  Fig.
\ref{DOS}b shows the corresponding results obtained by summing the terms
$(J^{xx}_{ij} - J^{yy}_{ij})$ over all lattice sites $j$ within a sphere of
radius $2.9\;a$ with $i$ taken within
the layers $1-8$ in the (Ga,Mn)As film. As one can see, the anisotropy of the
exchange interaction gives indeed a substantial contribution to the anisotropy
energy $E_{[1{1}0]} - E_{[1\overline{1}0]}$ for the layers 3 to 8. For
the first two film layers. i.e. surface and subsurface layers the two curves
strongly deviate reflecting the dominating on-site contribution to the MCA
\cite{USPW03,MBM+09}. 

In summary our  results show  that the tetragonal distortion
(caused by a compressive strain due to lattice mismatch of (Ga,Mn)As and GaAs
lattices) is responsible for the bi-axial in-plane anisotropy that is in line
with the interpretation given in previous investigations.  A strong uniaxial in-plane
MCA was found in (Ga,Mn)As film in the area adjacent to the surface or an
interface. We conclude that this is a result of the slow decay of the surface
potential gradient due to the small amount of free charge carriers. The
contribution to the uniaxial in-plane anisotropy decays rather slowly into the
bulk and is not restricted to only a few surface layers.  Moreover, a significant
contribution responsible for the MCA in the films is caused by the anisotropic
Mn-Mn exchange interactions mediated by holes in the valence band of (Ga,Mn)As.

\subsection{Acknowledgements}  

Financial support by the DFG through the SFB 689 is gratefully
acknowledged.


\begin{thebibliography}{26}%
\makeatletter
\providecommand \@ifxundefined [1]{%
 \@ifx{#1\undefined}
}%
\providecommand \@ifnum [1]{%
 \ifnum #1\expandafter \@firstoftwo
 \else \expandafter \@secondoftwo
 \fi
}%
\providecommand \@ifx [1]{%
 \ifx #1\expandafter \@firstoftwo
 \else \expandafter \@secondoftwo
 \fi
}%
\providecommand \natexlab [1]{#1}%
\providecommand \enquote  [1]{``#1''}%
\providecommand \bibnamefont  [1]{#1}%
\providecommand \bibfnamefont [1]{#1}%
\providecommand \citenamefont [1]{#1}%
\providecommand \href@noop [0]{\@secondoftwo}%
\providecommand \href [0]{\begingroup \@sanitize@url \@href}%
\providecommand \@href[1]{\@@startlink{#1}\@@href}%
\providecommand \@@href[1]{\endgroup#1\@@endlink}%
\providecommand \@sanitize@url [0]{\catcode `\\12\catcode `\$12\catcode
  `\&12\catcode `\#12\catcode `\^12\catcode `\_12\catcode `\%12\relax}%
\providecommand \@@startlink[1]{}%
\providecommand \@@endlink[0]{}%
\providecommand \url  [0]{\begingroup\@sanitize@url \@url }%
\providecommand \@url [1]{\endgroup\@href {#1}{\urlprefix }}%
\providecommand \urlprefix  [0]{URL }%
\providecommand \Eprint [0]{\href }%
\providecommand \doibase [0]{http://dx.doi.org/}%
\providecommand \selectlanguage [0]{\@gobble}%
\providecommand \bibinfo  [0]{\@secondoftwo}%
\providecommand \bibfield  [0]{\@secondoftwo}%
\providecommand \translation [1]{[#1]}%
\providecommand \BibitemOpen [0]{}%
\providecommand \bibitemStop [0]{}%
\providecommand \bibitemNoStop [0]{.\EOS\space}%
\providecommand \EOS [0]{\spacefactor3000\relax}%
\providecommand \BibitemShut  [1]{\csname bibitem#1\endcsname}%
\let\auto@bib@innerbib\@empty
\bibitem [{\citenamefont {Jungwirth}\ \emph {et~al.}(2006)\citenamefont
  {Jungwirth}, \citenamefont {Sinova}, \citenamefont {Ma\v{s}ek}, \citenamefont
  {Ku\v{c}era},\ and\ \citenamefont {MacDonald}}]{JSM+06}%
  \BibitemOpen
  \bibfield  {author} {\bibinfo {author} {\bibfnamefont {T.}~\bibnamefont
  {Jungwirth}}, \bibinfo {author} {\bibfnamefont {J.}~\bibnamefont {Sinova}},
  \bibinfo {author} {\bibfnamefont {J.}~\bibnamefont {Ma\v{s}ek}}, \bibinfo
  {author} {\bibfnamefont {J.}~\bibnamefont {Ku\v{c}era}}, \ and\ \bibinfo
  {author} {\bibfnamefont {A.~H.}\ \bibnamefont {MacDonald}},\ }\href@noop {}
  {\bibfield  {journal} {\bibinfo  {journal} {Rev. Mod. Phys.}\ }\textbf
  {\bibinfo {volume} {78}},\ \bibinfo {pages} {809} (\bibinfo {year}
  {2006})}\BibitemShut {NoStop}%
\bibitem [{\citenamefont {Abolfath}\ \emph {et~al.}(2001)\citenamefont
  {Abolfath}, \citenamefont {Jungwirth}, \citenamefont {Brum},\ and\
  \citenamefont {MacDonald}}]{AJBM01}%
  \BibitemOpen
  \bibfield  {author} {\bibinfo {author} {\bibfnamefont {M.}~\bibnamefont
  {Abolfath}}, \bibinfo {author} {\bibfnamefont {T.}~\bibnamefont {Jungwirth}},
  \bibinfo {author} {\bibfnamefont {J.}~\bibnamefont {Brum}}, \ and\ \bibinfo
  {author} {\bibfnamefont {A.~H.}\ \bibnamefont {MacDonald}},\ }\href@noop {}
  {\bibfield  {journal} {\bibinfo  {journal} {Phys. Rev. B}\ }\textbf {\bibinfo
  {volume} {63}},\ \bibinfo {pages} {054418} (\bibinfo {year}
  {2001})}\BibitemShut {NoStop}%
\bibitem [{\citenamefont {Dietl}\ \emph {et~al.}(2001)\citenamefont {Dietl},
  \citenamefont {Ohno},\ and\ \citenamefont {Matsukura}}]{DOM01}%
  \BibitemOpen
  \bibfield  {author} {\bibinfo {author} {\bibfnamefont {T.}~\bibnamefont
  {Dietl}}, \bibinfo {author} {\bibfnamefont {H.}~\bibnamefont {Ohno}}, \ and\
  \bibinfo {author} {\bibfnamefont {F.}~\bibnamefont {Matsukura}},\ }\href
  {\doibase 10.1103/PhysRevB.63.195205} {\bibfield  {journal} {\bibinfo
  {journal} {Phys. Rev. B}\ }\textbf {\bibinfo {volume} {63}},\ \bibinfo
  {pages} {195205} (\bibinfo {year} {2001})}\BibitemShut {NoStop}%
\bibitem [{\citenamefont {Sawicki}(2004)}]{Sav04}%
  \BibitemOpen
  \bibfield  {author} {\bibinfo {author} {\bibfnamefont {M.}~\bibnamefont
  {Sawicki}},\ }\href@noop {} {\bibfield  {journal} {\bibinfo  {journal} {Acta
  Physica Polonica A}\ }\textbf {\bibinfo {volume} {106}},\ \bibinfo {pages}
  {119} (\bibinfo {year} {2004})}\BibitemShut {NoStop}%
\bibitem [{\citenamefont {Dietl}(2001)}]{Die01}%
  \BibitemOpen
  \bibfield  {author} {\bibinfo {author} {\bibfnamefont {T.}~\bibnamefont
  {Dietl}},\ }\href
  {http://www.sciencedirect.com/science/article/pii/S1386947701000662}
  {\bibfield  {journal} {\bibinfo  {journal} {Physica E}\ }\textbf {\bibinfo
  {volume} {10}},\ \bibinfo {pages} {120} (\bibinfo {year} {2001})}\BibitemShut
  {NoStop}%
\bibitem [{\citenamefont {Sawicki}\ \emph {et~al.}(2003)\citenamefont
  {Sawicki}, \citenamefont {Matsukura}, \citenamefont {Dietl}, \citenamefont
  {Schott}, \citenamefont {Ruester}, \citenamefont {Schmidt}, \citenamefont
  {Molenkamp},\ and\ \citenamefont {Karczewski}}]{Sav03}%
  \BibitemOpen
  \bibfield  {author} {\bibinfo {author} {\bibfnamefont {M.}~\bibnamefont
  {Sawicki}}, \bibinfo {author} {\bibfnamefont {F.}~\bibnamefont {Matsukura}},
  \bibinfo {author} {\bibfnamefont {T.}~\bibnamefont {Dietl}}, \bibinfo
  {author} {\bibfnamefont {G.~M.}\ \bibnamefont {Schott}}, \bibinfo {author}
  {\bibfnamefont {C.}~\bibnamefont {Ruester}}, \bibinfo {author} {\bibfnamefont
  {G.}~\bibnamefont {Schmidt}}, \bibinfo {author} {\bibfnamefont {L.~W.}\
  \bibnamefont {Molenkamp}}, \ and\ \bibinfo {author} {\bibfnamefont
  {G.}~\bibnamefont {Karczewski}},\ }\href
  {http://dx.doi.org/10.1023/A:1023251710725} {\bibfield  {journal} {\bibinfo
  {journal} {Journal of Superconductivity}\ }\textbf {\bibinfo {volume} {16}},\
  \bibinfo {pages} {7} (\bibinfo {year} {2003})}\BibitemShut {NoStop}%
\bibitem [{\citenamefont {Liu}\ \emph {et~al.}(2003)\citenamefont {Liu},
  \citenamefont {Sasaki},\ and\ \citenamefont {Furdyna}}]{LSF03}%
  \BibitemOpen
  \bibfield  {author} {\bibinfo {author} {\bibfnamefont {X.}~\bibnamefont
  {Liu}}, \bibinfo {author} {\bibfnamefont {Y.}~\bibnamefont {Sasaki}}, \ and\
  \bibinfo {author} {\bibfnamefont {J.~K.}\ \bibnamefont {Furdyna}},\
  }\href@noop {} {\bibfield  {journal} {\bibinfo  {journal} {Phys. Rev. B}\
  }\textbf {\bibinfo {volume} {67}},\ \bibinfo {pages} {205204} (\bibinfo
  {year} {2003})}\BibitemShut {NoStop}%
\bibitem [{\citenamefont {Hamaya}\ \emph {et~al.}(2006)\citenamefont {Hamaya},
  \citenamefont {Watanabe}, \citenamefont {Taniyama}, \citenamefont {Oiwa},
  \citenamefont {Kitamoto},\ and\ \citenamefont {Yamazaki}}]{HWT+06}%
  \BibitemOpen
  \bibfield  {author} {\bibinfo {author} {\bibfnamefont {K.}~\bibnamefont
  {Hamaya}}, \bibinfo {author} {\bibfnamefont {T.}~\bibnamefont {Watanabe}},
  \bibinfo {author} {\bibfnamefont {T.}~\bibnamefont {Taniyama}}, \bibinfo
  {author} {\bibfnamefont {A.}~\bibnamefont {Oiwa}}, \bibinfo {author}
  {\bibfnamefont {Y.}~\bibnamefont {Kitamoto}}, \ and\ \bibinfo {author}
  {\bibfnamefont {Y.}~\bibnamefont {Yamazaki}},\ }\href@noop {} {\bibfield
  {journal} {\bibinfo  {journal} {Phys. Rev. B}\ }\textbf {\bibinfo {volume}
  {74}},\ \bibinfo {pages} {045201} (\bibinfo {year} {2006})}\BibitemShut
  {NoStop}%
\bibitem [{\citenamefont {Shin}\ \emph {et~al.}(2007)\citenamefont {Shin},
  \citenamefont {Chung}, \citenamefont {Lee}, \citenamefont {Liu},\ and\
  \citenamefont {Furdyna}}]{SCL+07}%
  \BibitemOpen
  \bibfield  {author} {\bibinfo {author} {\bibfnamefont {D.~Y.}\ \bibnamefont
  {Shin}}, \bibinfo {author} {\bibfnamefont {S.~J.}\ \bibnamefont {Chung}},
  \bibinfo {author} {\bibfnamefont {S.}~\bibnamefont {Lee}}, \bibinfo {author}
  {\bibfnamefont {X.}~\bibnamefont {Liu}}, \ and\ \bibinfo {author}
  {\bibfnamefont {J.~K.}\ \bibnamefont {Furdyna}},\ }\href@noop {} {\bibfield
  {journal} {\bibinfo  {journal} {Phys. Rev. B}\ }\textbf {\bibinfo {volume}
  {76}},\ \bibinfo {pages} {035327} (\bibinfo {year} {2007})}\BibitemShut
  {NoStop}%
\bibitem [{\citenamefont {Welp}\ \emph {et~al.}(2003)\citenamefont {Welp},
  \citenamefont {Vlasko-Vlasov}, \citenamefont {Liu}, \citenamefont {Furdyna},\
  and\ \citenamefont {Wojtowicz}}]{WVL+03}%
  \BibitemOpen
  \bibfield  {author} {\bibinfo {author} {\bibfnamefont {U.}~\bibnamefont
  {Welp}}, \bibinfo {author} {\bibfnamefont {V.~K.}\ \bibnamefont
  {Vlasko-Vlasov}}, \bibinfo {author} {\bibfnamefont {X.}~\bibnamefont {Liu}},
  \bibinfo {author} {\bibfnamefont {J.~K.}\ \bibnamefont {Furdyna}}, \ and\
  \bibinfo {author} {\bibfnamefont {T.}~\bibnamefont {Wojtowicz}},\ }\href@noop
  {} {\bibfield  {journal} {\bibinfo  {journal} {Phys. Rev. Lett.}\ }\textbf
  {\bibinfo {volume} {90}},\ \bibinfo {pages} {167206} (\bibinfo {year}
  {2003})}\BibitemShut {NoStop}%
\bibitem [{\citenamefont {Tang}\ \emph {et~al.}(2003)\citenamefont {Tang},
  \citenamefont {Kawakami}, \citenamefont {Awschalom},\ and\ \citenamefont
  {Roukes}}]{TKAR03}%
  \BibitemOpen
  \bibfield  {author} {\bibinfo {author} {\bibfnamefont {H.~X.}\ \bibnamefont
  {Tang}}, \bibinfo {author} {\bibfnamefont {R.~K.}\ \bibnamefont {Kawakami}},
  \bibinfo {author} {\bibfnamefont {D.~D.}\ \bibnamefont {Awschalom}}, \ and\
  \bibinfo {author} {\bibfnamefont {M.~L.}\ \bibnamefont {Roukes}},\ }\href
  {\doibase 10.1103/PhysRevLett.90.107201} {\bibfield  {journal} {\bibinfo
  {journal} {Phys. Rev. Lett.}\ }\textbf {\bibinfo {volume} {90}},\ \bibinfo
  {pages} {107201} (\bibinfo {year} {2003})}\BibitemShut {NoStop}%
\bibitem [{\citenamefont {Wang}\ \emph {et~al.}(2005)\citenamefont {Wang},
  \citenamefont {Sawicki}, \citenamefont {Edmonds}, \citenamefont {Campion},
  \citenamefont {Maat}, \citenamefont {Foxon}, \citenamefont {Gallagher},\ and\
  \citenamefont {Dietl}}]{WSE+05}%
  \BibitemOpen
  \bibfield  {author} {\bibinfo {author} {\bibfnamefont {K.-Y.}\ \bibnamefont
  {Wang}}, \bibinfo {author} {\bibfnamefont {M.}~\bibnamefont {Sawicki}},
  \bibinfo {author} {\bibfnamefont {K.~W.}\ \bibnamefont {Edmonds}}, \bibinfo
  {author} {\bibfnamefont {R.~P.}\ \bibnamefont {Campion}}, \bibinfo {author}
  {\bibfnamefont {S.}~\bibnamefont {Maat}}, \bibinfo {author} {\bibfnamefont
  {C.~T.}\ \bibnamefont {Foxon}}, \bibinfo {author} {\bibfnamefont {B.~L.}\
  \bibnamefont {Gallagher}}, \ and\ \bibinfo {author} {\bibfnamefont
  {T.}~\bibnamefont {Dietl}},\ }\href@noop {} {\bibfield  {journal} {\bibinfo
  {journal} {Phys. Rev. Lett.}\ }\textbf {\bibinfo {volume} {95}},\ \bibinfo
  {pages} {217204} (\bibinfo {year} {2005})}\BibitemShut {NoStop}%
\bibitem [{\citenamefont {Sugawara}\ \emph {et~al.}(2008)\citenamefont
  {Sugawara}, \citenamefont {Kasai}, \citenamefont {Tonomura}, \citenamefont
  {Brown}, \citenamefont {Campion}, \citenamefont {Edmonds}, \citenamefont
  {Gallagher}, \citenamefont {Zemen},\ and\ \citenamefont
  {Jungwirth}}]{SKT+08}%
  \BibitemOpen
  \bibfield  {author} {\bibinfo {author} {\bibfnamefont {A.}~\bibnamefont
  {Sugawara}}, \bibinfo {author} {\bibfnamefont {H.}~\bibnamefont {Kasai}},
  \bibinfo {author} {\bibfnamefont {A.}~\bibnamefont {Tonomura}}, \bibinfo
  {author} {\bibfnamefont {P.~D.}\ \bibnamefont {Brown}}, \bibinfo {author}
  {\bibfnamefont {R.~P.}\ \bibnamefont {Campion}}, \bibinfo {author}
  {\bibfnamefont {K.~W.}\ \bibnamefont {Edmonds}}, \bibinfo {author}
  {\bibfnamefont {B.~L.}\ \bibnamefont {Gallagher}}, \bibinfo {author}
  {\bibfnamefont {J.}~\bibnamefont {Zemen}}, \ and\ \bibinfo {author}
  {\bibfnamefont {T.}~\bibnamefont {Jungwirth}},\ }\href@noop {} {\bibfield
  {journal} {\bibinfo  {journal} {Phys. Rev. Lett.}\ }\textbf {\bibinfo
  {volume} {100}},\ \bibinfo {pages} {047202} (\bibinfo {year}
  {2008})}\BibitemShut {NoStop}%
\bibitem [{\citenamefont {Hrabovsky}\ \emph {et~al.}(2002)\citenamefont
  {Hrabovsky}, \citenamefont {Vanelle}, \citenamefont {Fert}, \citenamefont
  {Yee}, \citenamefont {Redoules}, \citenamefont {Sadowski}, \citenamefont
  {Kanski},\ and\ \citenamefont {Ilver}}]{HVF+02}%
  \BibitemOpen
  \bibfield  {author} {\bibinfo {author} {\bibfnamefont {D.}~\bibnamefont
  {Hrabovsky}}, \bibinfo {author} {\bibfnamefont {E.}~\bibnamefont {Vanelle}},
  \bibinfo {author} {\bibfnamefont {A.~R.}\ \bibnamefont {Fert}}, \bibinfo
  {author} {\bibfnamefont {D.~S.}\ \bibnamefont {Yee}}, \bibinfo {author}
  {\bibfnamefont {J.~P.}\ \bibnamefont {Redoules}}, \bibinfo {author}
  {\bibfnamefont {J.}~\bibnamefont {Sadowski}}, \bibinfo {author}
  {\bibfnamefont {J.}~\bibnamefont {Kanski}}, \ and\ \bibinfo {author}
  {\bibfnamefont {L.}~\bibnamefont {Ilver}},\ }\href@noop {} {\bibfield
  {journal} {\bibinfo  {journal} {Appl. Physics Lett.}\ }\textbf {\bibinfo
  {volume} {81}},\ \bibinfo {pages} {2806} (\bibinfo {year}
  {2002})}\BibitemShut {NoStop}%
\bibitem [{\citenamefont {Sawicki}\ \emph {et~al.}(2005)\citenamefont
  {Sawicki}, \citenamefont {Wang}, \citenamefont {Edmonds}, \citenamefont
  {Campion}, \citenamefont {Staddon}, \citenamefont {Farley}, \citenamefont
  {Foxon}, \citenamefont {Papis}, \citenamefont {Kaminska}, \citenamefont
  {Piotrowska}, \citenamefont {Dietl},\ and\ \citenamefont
  {Gallagher}}]{SWE+05}%
  \BibitemOpen
  \bibfield  {author} {\bibinfo {author} {\bibfnamefont {M.}~\bibnamefont
  {Sawicki}}, \bibinfo {author} {\bibfnamefont {K.-Y.}\ \bibnamefont {Wang}},
  \bibinfo {author} {\bibfnamefont {K.~W.}\ \bibnamefont {Edmonds}}, \bibinfo
  {author} {\bibfnamefont {R.~P.}\ \bibnamefont {Campion}}, \bibinfo {author}
  {\bibfnamefont {C.~R.}\ \bibnamefont {Staddon}}, \bibinfo {author}
  {\bibfnamefont {N.~R.~S.}\ \bibnamefont {Farley}}, \bibinfo {author}
  {\bibfnamefont {C.~T.}\ \bibnamefont {Foxon}}, \bibinfo {author}
  {\bibfnamefont {E.}~\bibnamefont {Papis}}, \bibinfo {author} {\bibfnamefont
  {E.}~\bibnamefont {Kaminska}}, \bibinfo {author} {\bibfnamefont
  {A.}~\bibnamefont {Piotrowska}}, \bibinfo {author} {\bibfnamefont
  {T.}~\bibnamefont {Dietl}}, \ and\ \bibinfo {author} {\bibfnamefont {B.~L.}\
  \bibnamefont {Gallagher}},\ }\href@noop {} {\bibfield  {journal} {\bibinfo
  {journal} {Phys. Rev. B}\ }\textbf {\bibinfo {volume} {71}},\ \bibinfo
  {pages} {121302(R)} (\bibinfo {year} {2005})}\BibitemShut {NoStop}%
\bibitem [{\citenamefont {Glunk}\ \emph {et~al.}(2009)\citenamefont {Glunk},
  \citenamefont {Daeubler}, \citenamefont {Dreher}, \citenamefont {Schwaiger},
  \citenamefont {Schoch}, \citenamefont {Sauer},\ and\ \citenamefont
  {Limmer}}]{GDD+09}%
  \BibitemOpen
  \bibfield  {author} {\bibinfo {author} {\bibfnamefont {M.}~\bibnamefont
  {Glunk}}, \bibinfo {author} {\bibfnamefont {J.}~\bibnamefont {Daeubler}},
  \bibinfo {author} {\bibfnamefont {L.}~\bibnamefont {Dreher}}, \bibinfo
  {author} {\bibfnamefont {S.}~\bibnamefont {Schwaiger}}, \bibinfo {author}
  {\bibfnamefont {W.}~\bibnamefont {Schoch}}, \bibinfo {author} {\bibfnamefont
  {R.}~\bibnamefont {Sauer}}, \ and\ \bibinfo {author} {\bibnamefont
  {Limmer}},\ }\href {\doibase 10.1103/PhysRevB.79.195206} {\bibfield
  {journal} {\bibinfo  {journal} {Phys. Rev. B}\ }\textbf {\bibinfo {volume}
  {79}},\ \bibinfo {pages} {195206} (\bibinfo {year} {2009})}\BibitemShut
  {NoStop}%
\bibitem [{\citenamefont {Zemen}\ \emph {et~al.}(2009)\citenamefont {Zemen},
  \citenamefont {Ku\v{c}era}, \citenamefont {Olejn\'ik},\ and\ \citenamefont
  {Jungwirth}}]{ZKOJ09}%
  \BibitemOpen
  \bibfield  {author} {\bibinfo {author} {\bibfnamefont {J.}~\bibnamefont
  {Zemen}}, \bibinfo {author} {\bibfnamefont {J.}~\bibnamefont {Ku\v{c}era}},
  \bibinfo {author} {\bibfnamefont {K.}~\bibnamefont {Olejn\'ik}}, \ and\
  \bibinfo {author} {\bibfnamefont {T.}~\bibnamefont {Jungwirth}},\ }\href@noop
  {} {\bibfield  {journal} {\bibinfo  {journal} {Phys. Rev. B}\ }\textbf
  {\bibinfo {volume} {80}},\ \bibinfo {pages} {155203} (\bibinfo {year}
  {2009})}\BibitemShut {NoStop}%
\bibitem [{\citenamefont {Kopeck\'y}\ \emph {et~al.}(2011)\citenamefont
  {Kopeck\'y}, \citenamefont {Kub}, \citenamefont {M\'aca}, \citenamefont
  {Ma\v{c}eck}, \citenamefont {Pacherov\'a}, \citenamefont {Rushforth},
  \citenamefont {Gallagher}, \citenamefont {Campion}, \citenamefont {Nov\'ak},\
  and\ \citenamefont {Jungwirth}}]{KKM+11}%
  \BibitemOpen
  \bibfield  {author} {\bibinfo {author} {\bibfnamefont {M.}~\bibnamefont
  {Kopeck\'y}}, \bibinfo {author} {\bibfnamefont {J.}~\bibnamefont {Kub}},
  \bibinfo {author} {\bibfnamefont {F.}~\bibnamefont {M\'aca}}, \bibinfo
  {author} {\bibfnamefont {J.}~\bibnamefont {Ma\v{c}eck}}, \bibinfo {author}
  {\bibfnamefont {O.}~\bibnamefont {Pacherov\'a}}, \bibinfo {author}
  {\bibfnamefont {A.~W.}\ \bibnamefont {Rushforth}}, \bibinfo {author}
  {\bibfnamefont {B.~L.}\ \bibnamefont {Gallagher}}, \bibinfo {author}
  {\bibfnamefont {R.~P.}\ \bibnamefont {Campion}}, \bibinfo {author}
  {\bibfnamefont {V.}~\bibnamefont {Nov\'ak}}, \ and\ \bibinfo {author}
  {\bibfnamefont {T.}~\bibnamefont {Jungwirth}},\ }\href {\doibase
  10.1103/PhysRevB.83.235324} {\bibfield  {journal} {\bibinfo  {journal} {Phys.
  Rev. B}\ }\textbf {\bibinfo {volume} {83}},\ \bibinfo {pages} {235324}
  (\bibinfo {year} {2011})}\BibitemShut {NoStop}%
\bibitem [{\citenamefont {Ebert}(2000)}]{Ebe00}%
  \BibitemOpen
  \bibfield  {author} {\bibinfo {author} {\bibfnamefont {H.}~\bibnamefont
  {Ebert}},\ }in\ \href@noop {} {\emph {\bibinfo {booktitle} {Electronic
  Structure and Physical Properties of Solids}}},\ \bibinfo {series} {Lecture
  Notes in Physics}, Vol.\ \bibinfo {volume} {535},\ \bibinfo {editor} {edited
  by\ \bibinfo {editor} {\bibfnamefont {H.}~\bibnamefont {Dreyss\'{e}}}}\
  (\bibinfo  {publisher} {Springer},\ \bibinfo {address} {Berlin},\ \bibinfo
  {year} {2000})\ p.\ \bibinfo {pages} {191}\BibitemShut {NoStop}%
\bibitem [{SPR(2009)}]{SPR-KKR5.4}%
  \BibitemOpen
  \href {http://olymp.cup.uni-muenchen.de/ak/ebert/SPRKKR} {}\bibinfo
  {howpublished} {{\em The Munich SPR-KKR package}, version 5.4, \newline
  \mbox{H.~Ebert~et~al.} \newline
  http://olymp.cup.uni-muenchen.de/ak/ebert/SPRKKR} (\bibinfo {year}
  {2009})\BibitemShut {NoStop}%
\bibitem [{\citenamefont {Staunton}\ \emph {et~al.}(2006)\citenamefont
  {Staunton}, \citenamefont {Szunyogh}, \citenamefont {Buruzs}, \citenamefont
  {Gyorffy}, \citenamefont {Ostanin},\ and\ \citenamefont {Udvardi}}]{SSB+06}%
  \BibitemOpen
  \bibfield  {author} {\bibinfo {author} {\bibfnamefont {J.~B.}\ \bibnamefont
  {Staunton}}, \bibinfo {author} {\bibfnamefont {L.}~\bibnamefont {Szunyogh}},
  \bibinfo {author} {\bibfnamefont {A.}~\bibnamefont {Buruzs}}, \bibinfo
  {author} {\bibfnamefont {B.~L.}\ \bibnamefont {Gyorffy}}, \bibinfo {author}
  {\bibfnamefont {S.}~\bibnamefont {Ostanin}}, \ and\ \bibinfo {author}
  {\bibfnamefont {L.}~\bibnamefont {Udvardi}},\ }\href {\doibase
  10.1103/PhysRevB.74.144411} {\bibfield  {journal} {\bibinfo  {journal} {Phys.
  Rev. B}\ }\textbf {\bibinfo {volume} {74}},\ \bibinfo {pages} {144411}
  (\bibinfo {year} {2006})}\BibitemShut {NoStop}%
\bibitem [{\citenamefont {Ebert}\ and\ \citenamefont
  {Mankovsky}(2009)}]{EM09a}%
  \BibitemOpen
  \bibfield  {author} {\bibinfo {author} {\bibfnamefont {H.}~\bibnamefont
  {Ebert}}\ and\ \bibinfo {author} {\bibfnamefont {S.}~\bibnamefont
  {Mankovsky}},\ }\href {\doibase 10.1103/PhysRevB.79.045209} {\bibfield
  {journal} {\bibinfo  {journal} {Phys. Rev. B}\ }\textbf {\bibinfo {volume}
  {79}},\ \bibinfo {pages} {045209} (\bibinfo {year} {2009})}\BibitemShut
  {NoStop}%
\bibitem [{\citenamefont {Udvardi}\ \emph {et~al.}(2003)\citenamefont
  {Udvardi}, \citenamefont {Szunyogh}, \citenamefont {Palot\'as},\ and\
  \citenamefont {Weinberger}}]{USPW03}%
  \BibitemOpen
  \bibfield  {author} {\bibinfo {author} {\bibfnamefont {L.}~\bibnamefont
  {Udvardi}}, \bibinfo {author} {\bibfnamefont {L.}~\bibnamefont {Szunyogh}},
  \bibinfo {author} {\bibfnamefont {K.}~\bibnamefont {Palot\'as}}, \ and\
  \bibinfo {author} {\bibfnamefont {P.}~\bibnamefont {Weinberger}},\ }\href
  {\doibase 10.1103/PhysRevB.68.104436} {\bibfield  {journal} {\bibinfo
  {journal} {Phys. Rev. B}\ }\textbf {\bibinfo {volume} {68}},\ \bibinfo
  {pages} {104436} (\bibinfo {year} {2003})}\BibitemShut {NoStop}%
\bibitem [{\citenamefont {Mankovsky}\ \emph {et~al.}(2009)\citenamefont
  {Mankovsky}, \citenamefont {Bornemann}, \citenamefont {Min\'ar},
  \citenamefont {Polesya}, \citenamefont {Ebert}, \citenamefont {Staunton},\
  and\ \citenamefont {Lichtenstein}}]{MBM+09}%
  \BibitemOpen
  \bibfield  {author} {\bibinfo {author} {\bibfnamefont {S.}~\bibnamefont
  {Mankovsky}}, \bibinfo {author} {\bibfnamefont {S.}~\bibnamefont
  {Bornemann}}, \bibinfo {author} {\bibfnamefont {J.}~\bibnamefont {Min\'ar}},
  \bibinfo {author} {\bibfnamefont {S.}~\bibnamefont {Polesya}}, \bibinfo
  {author} {\bibfnamefont {H.}~\bibnamefont {Ebert}}, \bibinfo {author}
  {\bibfnamefont {J.~B.}\ \bibnamefont {Staunton}}, \ and\ \bibinfo {author}
  {\bibfnamefont {A.~I.}\ \bibnamefont {Lichtenstein}},\ }\href {\doibase
  10.1103/PhysRevB.80.014422} {\bibfield  {journal} {\bibinfo  {journal} {Phys.
  Rev. B}\ }\textbf {\bibinfo {volume} {80}},\ \bibinfo {pages} {014422}
  (\bibinfo {year} {2009})}\BibitemShut {NoStop}%
\bibitem [{\citenamefont {Daeubler}\ \emph {et~al.}(2008)\citenamefont
  {Daeubler}, \citenamefont {Schwaiger}, \citenamefont {Glunk}, \citenamefont
  {Tabor}, \citenamefont {Schoch}, \citenamefont {Sauer},\ and\ \citenamefont
  {Limmer}}]{DSG+08}%
  \BibitemOpen
  \bibfield  {author} {\bibinfo {author} {\bibfnamefont {J.}~\bibnamefont
  {Daeubler}}, \bibinfo {author} {\bibfnamefont {S.}~\bibnamefont {Schwaiger}},
  \bibinfo {author} {\bibfnamefont {M.}~\bibnamefont {Glunk}}, \bibinfo
  {author} {\bibfnamefont {M.}~\bibnamefont {Tabor}}, \bibinfo {author}
  {\bibfnamefont {W.}~\bibnamefont {Schoch}}, \bibinfo {author} {\bibfnamefont
  {R.}~\bibnamefont {Sauer}}, \ and\ \bibinfo {author} {\bibfnamefont
  {W.}~\bibnamefont {Limmer}},\ }\href
  {http://www.sciencedirect.com/science/article/pii/S1386947707003530}
  {\bibfield  {journal} {\bibinfo  {journal} {Physica E}\ }\textbf {\bibinfo
  {volume} {40}},\ \bibinfo {pages} {1876} (\bibinfo {year}
  {2008})}\BibitemShut {NoStop}%
\bibitem [{\citenamefont {Ko\v{s}uth}(2007)}]{Kos07}%
  \BibitemOpen
  \bibfield  {author} {\bibinfo {author} {\bibfnamefont {M.}~\bibnamefont
  {Ko\v{s}uth}},\ }\emph {\bibinfo {title} {Magnetic properties of transition
  metal surfaces and GaAs/Fe heterogeneous systems}},\ \href@noop {} {Ph.D.
  thesis},\ \bibinfo  {school} {University of Munich} (\bibinfo {year}
  {2007})\BibitemShut {NoStop}%
\end{thebibliography}

%

\end{document}